\documentstyle[aps,aps10]{revtex}
\begin{document}
\def\beq{\begin{equation}}
\def\eq{\end{equation}}
\def\etal{{\it et al.}}
\def\pC{$pC^{12}$}
\title{Subthreshold $K^+$ Meson Production in Proton--Nucleus Reactions}
\author{S.V. Efremov\\
{\it Bonner Nuclear Laboratory, Rice University, P.O. Box 1892,} \\
{\it Houston, TX 77251-1892, USA}
\and \\E.Ya. Paryev\\
{\it Institute for Nuclear Research, Russian Academy of Sciences,}\\
{\it Moscow 117312, Russia}}
\maketitle
\begin{abstract}
	  The inclusive $K^+$ mesons production in proton--nucleus collisions 
        in the near threshold and subthreshold energy regimes is analyzed 
        with respect to the one--step ($pN \to K^+YN$, $Y=\Lambda,\Sigma$) and 
        the two--step ($pN \to NN\pi, NN2\pi; ~ \pi N \to K^+Y$) incoherent 
	  production processes on the basis of an appropriate new folding model, 
	  which takes properly into account 
        the struck target nucleon removal energy and momentum
        distribution (nucleon spectral function), extracted from recent
        quasielastic electron scattering experiments and from many--body
        calculations with realistic models of the $NN$ interaction. Comparison
        of the model calculations of the $K^+$ double differential cross
        sections for the reaction $p+C^{12}$ at 1.2, 1.5 and
        $2.5~GeV$ beam energies with the existing experimental data from the
        SATURNE experiment is given, illustrating both the relative role
        of the primary and secondary production channels at considered
        incident energies and those features of the cross sections which are
        sensitive to the high momentum and high removal energy part of the
        nucleon spectral function that is governed by nucleon--nucleon
        short--range and tensor correlations. In--medium modifications 
        of the available for pion and hyperon production invariant energies 
        squared due to the respective optical potentials are needed 
        to account for the experimental data on $K^+$ production in the energy 
        region under consideration. 
\end{abstract}

\section*{Introduction}
	Extensive investigations of the production of $K^+$ mesons in
	proton--nucleus reactions [1--9] at 
	incident energies lower than the free nucleon-nucleon threshold
	have been carried out in the last years. The study of inclusive
        and exclusive subthreshold $K^+$ production in $pA$ interactions
        is planned in the near future at the accelerators
        COSY--J$\ddot{\rm u}$lich [10] and CELSIUS [11]. 
        Because of the high $K^+$ production threshold ($1.58~GeV$) 
        in the nucleon-nucleon collision and the
	rather weak $K^+$ rescattering in the surrounding medium compared to 
	the pions, etas, antiprotons and antikaons, from these studies
        one hopes to extract information about both the intrinsic properties 
        of target nuclei (such as Fermi motion, high momentum components 
        of the nuclear wave function, clusters of nucleons or quarks) and 
	reaction mechanism, in--medium properties of hadrons. 
	However, in spite of large efforts, subthreshold kaon production 
        in proton--nucleus reactions is still far from being 
	fully understood. Thus, measured total $K^+$ production cross 
        sections [1] in the range of proton energies of 800--1000 $MeV$ were 
        described in the framework of the respective folding models based on 
        both the direct mechanism [2,6] of $K^+$ production 
        ($pN\rightarrow K^+\Lambda N$) and on the two--step mechanism
        [3--5] associated with the production of kaons by intermediate pions
        ($pN_1\rightarrow \pi NN,\, \pi N_2\rightarrow K^+\Lambda$) using
        different parametrizations for internal nucleon momentum
        distribution [6, 12, 13]. The same experimental data could be well 
        explained also in the modified phase space model [7]. 
        In the above folding
        models only the internal nucleon momentum distribution has been used
        and the off--shell propagation of the struck target nucleon has been
        neglected or has been taken into account the most crudely, but it 
        could be significant in processes that are limited by phase space 
        such as the threshold heavy mesons production. As is well known 
        [13--21], the off--shell behaviour of a bound nucleon is described
        by the nucleon spectral function $P({\bf p}_t,E)$, which represents
        the probability to find in the nucleus a nucleon with momentum
        ${\bf p}_t$ and removal (binding) energy E and contains all the
        information on the structure of a target nucleus. The knowledge
        of the spectral function $P({\bf p}_t,E)$ is needed for calculations
        of cross sections of various kinds of nuclear reactions. In particular,
        it has been widely used earlier for the analysis of inclusive and
        exclusive quasielastic electron scattering by nuclei [13--26]. It was
        found that even at very high momentum and energy transfer
        in this scattering, the target nucleus cannot be simply described as a
        collection of A on--shell nucleons subject only to Fermi motion, but
        the full nucleon momentum and binding energy distribution has to be
        considered. Only recently [8, 9] the binding energy of the struck
        target nucleon has been properly taken into account in calculating
        the subthreshold kaon production in $pA$ collisions. Debowski et al.
        [8] analyzed their own double differential cross sections data
        for the $K^+$ production in $p+C$ reactions at 1.2, 1.5 and $2.5~GeV$
        beam energies in the framework of the first chance collision model
        with the nucleon spectral function taken from [20, 27]. It was shown
        that at subthreshold incident energies the $K^+$ production cross
        sections are underestimated significantly by calculations assuming
        only first chance collisions. The same nucleon spectral function has
        been employed by Sibirtsev et al. [9] for the description of the
        measured [1] total cross sections for $K^+$ production from
        $p+C$ collisions in the framework of the two--step model. The claims 
        have been advanced in [9] that within the spectral function approach
        the two--step production mechanism with an intermediate pion also
        dominates at subthreshold energies as in the folding models [3--5],
        but the calculated total cross sections for $K^+$ production are
        underestimated below about 920 $MeV$ in contrast to the folding model
        where a better description of the experimental data is achieved
        as well as the contribution to the calculated total cross sections
        from the high momentum and high removal energy part (correlated part)
        of the nucleon spectral function is small compared to that calculated
        with the single--particle (uncorrelated) part of the one. However,
        in order to fully clarify the role played by nucleon--nucleon
        correlations induced by realistic interactions in the subthreshold
        kaon production in $pA$ collisions as well as to get a deeper insight
        into the relative role of the primary and secondary reaction channels,
        it is necessary to properly analyze within the same approach,
        based on nucleon spectral function, the existing experimental data both
        on total [1] and differential [8] kaon production cross sections.

        In this paper we have performed such description of the total and
        differential $K^+$ production cross sections from $pC$ reactions
        in the near threshold and subthreshold energy regimes using the
        appropriate new folding model for primary and secondary production
        processes, which takes into account the struck target nucleon
        removal energy and momentum distribution.    

\section{ Direct $K^+$ Production Process} 

	Apart from participation in the elastic scattering an incident proton
	can produce a $K^+$ directly in the first inelastic
	$pN$ collision due to nucleon Fermi motion. 
	Since we are interested in a few $GeV$ region (up to $2.5~GeV$), 
	we have taken into account [28]
	the following elementary processes which have the lowest 
	free production thresholds: 
\beq
p+N \to K^++\Lambda+N,
\eq
\beq
p+N \to K^++\Sigma+N.
\eq
	Neglecting the kaon rescatterings in the nuclear medium [29],
	we can represent the invariant inclusive cross section of $K^+$
	production on nuclei by the initial proton with momentum
	${\bf p}_0$ as follows [13, 23, 29]:	 

\beq
E_{K^+}\frac{d\sigma_{pA\to K^+X}^{(prim)}({\bf p}_0)}
{d{\bf p}_{K^+}}=
I_V[A]\times
\eq
$$
\left\{\left<E_{K^+}\frac{d\sigma_{pN\to K^+\Lambda N}({\bf p}_0,{\bf p}_{K^+})}
{d{\bf p}_{K^+}}\right>+
\left<E_{K^+}\frac{d\sigma_{pN\to K^+\Sigma N}({\bf p}_0,{\bf p}_{K^+})}
{d{\bf p}_{K^+}}\right>\right\},
$$
where
\beq
\left<E_{K^+}\frac{d\sigma_{pN\to K^+YN}({\bf p}_0,{\bf p}_{K^+})}
{d{\bf p}_{K^+}}\right>=
\int\int 
P({\bf p}_t,E)d{\bf p}_tdE\times
\eq
$$
\left[E_{K^+}
\frac
{d\sigma_{pN\to K^+YN}(\sqrt{s}, {\bf p}_{K^+})}
{d{\bf p}_{K^+}}\right].
$$
Here, 
$E_{K^+}d\sigma_{pN\to K^+YN}(\sqrt{s},{\bf p}_{K^+})/{d{\bf p}_{K^+}}$
are the free invariant inclusive cross sections for  $K^+$ 
production in reactions (1)--(2); 
$P({\bf p}_t,E)$ is the nucleon spectral function normalized to unity; 
${\bf p}_t$ and E are
the internal momentum and removal energy of the struck target nucleon 
just before the collision; 
${\bf p}_{K^+}$ and $E_{K^+}$ are the momentum and total energy of 
a $K^+$ meson, respectively;
$I_V[A]$ is the effective number of nucleons for the $pN\to K^+YN$ reaction
on nuclei
\footnote
{Calculations show that in the case of Gaussian nuclear density for $C^{12}$
target nucleus the effective number of nucleons is equal to 6.9.}; 
$s$ is the $pN$ center--of--mass energy squared. 
The expression for $s$ is: 
\beq
  s=(E_0+E_t)^2-({\bf p}_0+{\bf p}_t)^2,
\eq
  where $E_0$ and $E_t$ are the projectile's total energy, given by $E_0=
  \sqrt{p_{0}^{2}+m_{N}^{2}}$ ($m_{N}$ is the rest mass of a nucleon), 
  and the struck target nucleon total energy, respectively. 
  Taking into account the recoil and excitation energies of the residual 
  $(A-1)$ system, one has [23]:
\beq
   E_t=M_A-\sqrt{(-{\bf p}_t)^2+(M_{A}-m_{N}+E)^{2}},
\eq
 where $M_A$ is the mass of the initial target nucleus.
 It is easily seen that
 in this case the struck target nucleon is off--shell. 
 In Eq. (3) it is assumed that the $K^+$ meson production cross sections in
 $pp$-- and $pn$--interactions are the same [8,28] as well as it is disregarded
 any difference between the proton and the neutron spectral functions [18].
 In our approach the invariant inclusive cross sections for $K^+$ production
 in the elementary processes (1)--(2) have been described by the three--body
 phase space calculations normalized to the corresponding total cross
 sections [28].

 The nucleon spectral function, $P({\bf p}_t,E)$, is a crucial point in the
 evaluation of the subthreshold production of any particles on a nuclear
 target. When ground--state $NN$ correlations, which are generated by the
 short range and tensor parts of realistic $NN$ interaction, are considered,
 the spectral function $P({\bf p}_t,E)$ can be represented in the following
 form [16--21]:    
\beq
  P({\bf p}_t,E)=P_{0}({\bf p}_t,E)+P_{1}({\bf p}_t,E), 
\eq
 where $P_{0}$ includes the ground and one--hole states of the residual
 $(A-1)$ nucleon system and $P_{1}$ more complex configurations (mainly
 1p--2h states) that are arise from 2p--2h excited states generated in the
 ground state of the target nucleus by $NN$ correlations. In the absence of NN
 correlations (within the Hartree--Fock approximation) function 
 $P_{1}({\bf p}_t,E)=0$ and the Hartree--Fock spectral function [16--21]
 is recovered. Thus, the $NN$ correlations deplete ($20\pm5\%$, 
 [16, 17, 21, 26, 30])
 states below the Fermi sea and populate states outside the Fermi sea.  
 For $K^+$ production calculations in the case of $C^{12}$ target nucleus
 reported here we have employed for the single--particle part 
 $P_{0}({\bf p}_t,E)$ of the nucleon spectral function the following
 relation [16--18]: $P_{0}({\bf p}_t,E)=0.8 P^{(SM)}({\bf p}_t,E)$ with
 $P^{(SM)}({\bf p}_t,E)$ being the harmonic oscillator spectral function
 (see, e.g., [24]) in which the s-- and p--shell nucleon momentum distributions
 were taken from [29] and binding energies of 34 and 16 $MeV$ [24] for the
 s and p shells, respectively, were used. 

 Results of the many--body calculations with realistic models of the NN
 interaction in few--body systems, complex nuclei and nuclear matter
 [16, 17, 19--23] as well as experimental data for nucleon momentum distribution
 $n({\bf p}_t)=\int P({\bf p}_t,E)dE$ obtained by the y--scaling
 analysis [18] show that: 
\begin{itemize}
\item[$i)$] the momentum distribution $n({\bf p}_t)$ for
 $p_t > 2 fm^{-1}$ is a several orders of magnitude larger than the
 predictions from mean--field calculations and, moreover, the high momentum
 behaviour of $n({\bf p}_t)$ (at $p_t\ge 2 fm^{-1}$) is similar for nuclei
 with mass number $A \ge 2$; 
\item[$ii)$] the behaviour of the nucleon spectral function
 at high values of $p_t$ and E is almost entirely governed by
 $P_{1}({\bf p}_t,E)$ as well as there is a strong coupling between the high
 momentum components and the high values of the removal energy
 ($E \sim p_{t}^{2}/2m_{N}$) of a nucleon embedded in the nuclear medium.
\end{itemize} 
 Starting from such observation and analyzing the perturbative expansion of
 the $NN$ interaction and the momentum distribution for potentials decreasing
 at large values of $p_{t}$ as powers of $p_{t}$, Frankfurt and Strikman
 have argued [31] that the structure of the nucleon spectral function at
 high values of nucleon momentum and removal energy should be generated
 mainly by the ground--state two--nucleon configurations with large relative
 ($\ge 1.5~fm^{-1}$), but low ($< 1.5~fm^{-1}$) center--of--mass momenta.
 Basing on this assumption, the convolution model for the correlated part
 $P_{1}({\bf p}_t,E)$ of the nucleon spectral function in the case of any
 value of the mass number A has been proposed [19, 21] according to which
 function $P_{1}({\bf p}_t,E)$ is expressed as a convolution integral of
 the momentum distributions describing the relative and center--of--mass
 motion of a correlated $NN$ pair embedded in the nuclear medium. It was shown
 [19, 21] that this model satisfactorily reproduces the existing spectral
 functions of $He^{3}$, $He^{4}$ and nuclear matter obtained by means of
 many--body calculations with realistic models of the $NN$ interaction.
 An inspection of the convolution formula (53) from [21] for the spectral
 function $P_{1}({\bf p}_t,E)$ leads to the simple analytical expression
 for the $P_{1}({\bf p}_t,E)$ proposed in [17] (formula (7)). This expression
 for the quantity $P_{1}({\bf p}_t,E)$ (in which the correlated part 
 $n_{1}({\bf p}_t)$ of the $n({\bf p}_t)$
 in the case of $C^{12}$ target nucleus was taken from [16], the peak position
 and the FWHM were taken from [21]) was used in our calculations of $K^+$
 production in $pC$ collisions.

	Let us consider now the two--step $K^+$ production mechanism.

\section{ Two--Step $K^+$ Production Process}

	Kinematical considerations show that in the bombarding energy range
of our interest ($\le 2.5~GeV$) the following two--step production
process may not only contribute to the $K^+$ production in $pA$ interactions
but even dominate [1, 3--5] at subthreshold energies ($\le 1.5~GeV$).
An incident proton can produce in the first inelastic collision with an 
intranuclear nucleon also a pion through the elementary reactions:
\beq
p+N_{1}\to N+N+\pi,
\eq
\beq
p+N_{1}\to N+N+2\pi.
\eq
We remind that the free threshold energies for these reactions respectively
are 0.29 and $0.60~GeV$. Then the intermediate pion, which is assumed to be
on--shell, produces the kaon on a nucleon of the target nucleus via 
the elementary subprocesses with the lowest free production thresholds 
(respectively, 0.76 and $0.89~GeV$):
\beq
\pi + N_2\to K^++\Lambda,
\eq 
\beq
\pi + N_2\to K^++\Sigma,
\eq
provided that these subprocesses are energetically possible. 
In order to calculate the $K^+$ production
cross section for $pA$ reactions from the secondary pion induced 
reaction channels (10), (11) we fold the momentum--energy--averaged 
inclusive differential 
cross section for pion production in the reactions (8), (9)
(denoted by 
$<d\sigma_{p N\to \pi X}({\bf p}_0,{\bf p}_{\pi})/
d{\bf p}_{\pi}>$) 
with the momentum--energy--averaged inclusive invariant 
differential cross section for $K^+$ production in these channels 
(denoted by $<E_{K^+}d\sigma_{\pi N\to K^+X}
({\bf p}_{\pi},{\bf p}_{K^+})/d{\bf p}_{K^+}>$)
and the effective number of $NN$ pairs per unit of square
(denoted via 
$I_V[A,\sigma_{pN}^{in}(p_0),\sigma_{\pi N}^{tot}(p_{\pi}),\vartheta_{\pi}]$), 
i.e. (see, also, [29]):
\beq
E_{K^+}\frac
{d\sigma_{pA\to K^+X}^{(sec)}}
{d{\bf p}_{K^+}}=
\int d{\bf p}_{\pi}
I_V[A,\sigma_{pN}^{in}(p_0),
\sigma_{\pi N}^{tot}(p_{\pi}),\vartheta_{\pi}]\times
\eq
$$\times
\left<\frac{d\sigma_{pN\to \pi X}
({\bf p}_0,{\bf p}_{\pi})}{d{\bf p}_{\pi}}\right>
\left<E_{K^+}\frac{d\sigma_{\pi N\to K^+X}
({\bf p}_{\pi},{\bf p}_{K^+})}{d{\bf p}_{K^+}}\right>,
$$
Here, ${\bf p}_{\pi}$ is the momentum of a pion. The quantity
$I_V[A,\sigma_{pN}^{in}(p_0),\sigma_{\pi N}^{tot}(p_{\pi}),\vartheta_{\pi}]$)
in Eq. (12) is determined by the formula (39) from [29] in which one has to
make the following substitutions: $NZ \to A^{2}$, $\eta \to \pi$, and
$\sigma_{\pi^+N}^{tot} \to \sigma_{pN}^{in}$. It should be noted that the
symbol $\int d{\bf p}_{\pi}$ in Eq. (12) stands for summation over the pion
isospin and integration over the pion momentum space.
   
In our method the differential cross sections for pion production in the
elementary processes (8), (9) have been described by the three-- and four--body
phase space calculations normalized to the respective total cross 
sections [32]. The invariant inclusive cross sections for $K^+$ production
in the elementary reactions (10), (11) have been described by the two--body
phase space calculations [29] normalized to the corresponding total cross
sections 
$\sigma_{\pi N\to K^+\Lambda}$ [33] and $\sigma_{\pi N\to K^+\Sigma}$ [34].
We have taken into account in the calculation of the $K^+$ production cross
section (12) from the two--step processes (8)--(11) also the following
medium effects
\footnote
{We have neglected the analogous medium effects in calculating the $K^+$
production cross section (3) in the one--step processes (1), (2), since
the contributions from them to the invariant energy squared $s$ compensate
each other.}: 
the influence of the average momentum--dependent optical potential on the
incoming proton (denoted by $V_{0}$), on produced in the reactions (8), (9)
nucleons (denoted by $V_{N}$), (10), (11) hyperons (denoted by $V_{Y}$)
\footnote
{The pion and kaon potentials in a nuclear medium were set to zero [35].}
. The available in this case for the pion and hyperon production invariant
energies squared of the pN-- (s) and $\pi N$--systems ($s_{1}$) can be
written as follows (cf. with Eq. (5)):
\beq
  s=(E_0^{'}-2V_N+E_t)^2-({\bf p}_0^{'}+{\bf p}_t)^2,
\eq
\beq
  s_1=(E_{\pi}-V_Y+E_t)^2-({\bf p}_{\pi}+{\bf p}_t)^2,
\eq
where $E_0^{'}$, ${\bf p}_0^{'}$ are the in--medium total energy and
momentum of the initial proton ($E_0^{'}=E_0-V_0$, 
${p_0^{'}}^2=p_0^{2}-2m_{N}V_0-V_0^{2}$); $E_{\pi}$ is the total energy of
the intermediate pion.

	Now, let us discuss the results of our calculations in the 
framework of approach outlined above.

\section{Results and Discussion}

The comparison of the results of our calculations by (3)--(7), (12)--(14)
with the experimental data [8] for the double differential cross sections
for the production of $K^+$ mesons at an angle of 40$^{0}$ in the interaction
of protons with energies of 1.2, 1.5 and $2.5~GeV$ with $C^{12}$ nuclei is
given in Figs.1--3. It is seen that: 
\begin{itemize}
\item[$1)$] the two--step $K^+$ production
mechanism clearly dominates at subthreshold beam energies in accordance with
earlier calculations [3--5] based on the folding model with one--dimensional
internal nucleon momentum distributions and it is of minor importance at
$2.5~GeV$ proton beam energy at which $K^+$ mesons are produced almost entirely
in first chance collisions (1), (2) in line with our earlier calculations [28];
\item[$2)$] the main contribution to the $K^+$ production in the two--step 
reaction
channels (8)--(11) both at subthreshold and above the free $NN$ threshold
incident energies considered here comes from the use only of the uncorrelated
part $P_{0}({\bf p}_t,E)$ of the nucleon spectral function in the calculation
of the corresponding momentum--energy--averaged differential cross sections
for pion and kaon production; 
\item[$3)$] the $K^+$ production cross section in the
two--step reaction channels (8)--(11) coming from the use of the uncorrelated
and correlated parts of the nucleon spectral function in the calculation of
momentum--energy--averaged differential cross sections for pion and kaon
production, respectively, is approximately the same as that obtained from 
the use of the correlated and uncorrelated parts of one in the analogous
calculation; 
\item[$4)$] the contribution to the $K^+$ production in the same reaction
channels only from the correlated part $P_{1}({\bf p}_t,E)$ of the nucleon
spectral function  is an order of magnitude smaller than the contributions
from the use of the uncorrelated (correlated) and correlated (uncorrelated)
parts of one in calculating the momentum--energy--averaged differential cross
sections for pion and kaon production, respectively; 
\item[$5)$] the contribution to
the $K^+$ production from the secondary reaction channels (10) and (11) with
a $\Lambda$ and a $\Sigma$ particles in the final states are comparable at
$2.5~GeV$ beam energy, whereas at subthreshold energies the secondary
production process (10) is more important than the (11) one; 
\item[$6)$] our full
calculations (the sum of results obtained both for the one--step (1), (2) and
two--step (8)--(11) reaction channels, solid lines) reproduce the experimental
data only if we take into account in the two--step production processes 
(8)--(11) the in--medium modifications according to Eqs.(13), (14) of the
available for pion and hyperon production invariant energies squared of the
pN-- and $\pi N$--systems.
\end{itemize}
  Finally, it should be noted that our model calculations for the two--step
production processes (8)--(11)  with the use only of uncorrelated part of
the nucleon spectral function as well as the same in--medium modifications
of the available for pion and hyperon production invariant energies squared
as in above reproduce also the measured [1] total $K^+$ production cross 
sections in $pC$ collisions at proton energies above about 900 $MeV$ but
underestimate the data at lower bombarding energies, what indicates the need
for other reaction channels or multinucleon correlations to explain these
data close to the absolute production threshold.
    
  Taking into account the considered above, we conclude that the high
momentum and high removal energy part of the nucleon spectral function which
is generated by ground--state two--nucleon short--range correlations apparently
cannot be studied in the inclusive subthreshold kaon production in 
$pA$ reactions. As it was shown in [29], the $\pi^++A \to K^++X$ reaction in
the subthreshold regime seems to be quite promising for this aim.

\section{Summary}

In this paper we have calculated the total and differential cross sections
for $K^+$ production from $p+C^{12}$ reactions in the near threshold and
subthreshold energy regimes by considering incoherent primary proton--nucleon 
and secondary pion--nucleon production processes within the framework of an 
appropriate folding model, which takes properly into account the struck target 
nucleon momentum and removal energy distribution. The comparison of the 
results of our calculations with the existing experimental data [1, 8] was made.
It was found that the in--medium modifications of the available for pion and
hyperon production invariant energies squared due to the respective optical
potentials are needed to account for the experimental data both from
Koptev et al. [1] and from the SATURNE/GSI collaboration [8]. It was shown also
that the two--step $K^+$ production mechanism clearly
dominates at subthreshold energies in line with earlier calculations [3--5]
based on the use of the one--dimensional momentum distributions of nucleons
in the target nucleus. Our investigations indicate that the main contribution
to $K^+$ production in the secondary reaction channels comes from the use only
of the uncorrelated part of the nucleon spectral function. Therefore, inclusive 
subthreshold kaon production in $pA$ collisions does not provide an information
on the high momentum components within target nucleus.

\end{document}